\documentclass[final,5p,times,twocolumn]{elsarticle}

\usepackage{amssymb}
\usepackage{amsmath}
\usepackage{lipsum}
\usepackage[T1]{fontenc}

\journal{Physics Letters B}

\newcommand{\pr}{p_R}
\newcommand{\cum}[2]{\left \langle {#1}^{#2} \right \rangle_{\!c}}
\newcommand{\delN}{\Delta N}
\newcommand{\netp}{{p\!-\!\bar p}}

\begin{document}

\begin{frontmatter}



\title{Excited cluster states: A new source for proton number fluctuations \\ in the high baryon density regime}


\author[FJFI,UMB]{\protect{Boris Tom\'a\v{s}ik}}
\author[GU,HFHF]{Marcus Bleicher}
\affiliation[FJFI]{organization={Fakulta jadern\'a a fyzik\'aln\v{e} in\v{z}en\'yrsk\'a, \v{C}esk\'e vysok\'e u\v{c}en\'i technick\'e v Praze},
            addressline={{B\v{r}ehov\'a 7}}, 
            city={Praha 1},
            postcode={11519}, 
            country={Czech Republic}}
\affiliation[UMB]{organization={Fakulta pr\'irodn\'ych vied, Univerzita Mateja Bela},
            addressline={Tajovsk\'eho 40}, 
            city={Bansk\'a Bystrica},
            postcode={97401}, 
            country={Slovakia}}      
\affiliation[GU]{organization={Institut f\"{u}r Theoretische Physik, Goethe Universit\"{a}t Frankfurt},
            addressline={Max-von-Laue-Str.~1}, 
            city={Frankfurt am Main},
            postcode={60438}, 
            country={Germany}}
\affiliation[HFHF]{organization={Helmholtz Research Academy Hesse for FAIR (HFHF), GSI Helmholtzzentrum f\"ur Schwerionenforschung GmbH, Campus Frankfurt},
            addressline={Max-von-Laue-Str.~12}, 
            city={Frankfurt am Main},
            postcode={60438}, 
            country={Germany}}

\begin{abstract}
We calculate the contribution of the decay products of excited nuclear cluster states to the event-by-event fluctuations of protons in the energy range from $\sqrt{s_{NN}}=2-5$~GeV within the statistical model. We find that the inclusion of the excited nuclear clusters yields corrections to all cumulant ratios, ranging from 1\% for ratio of second to first-order cumulant to 100\% for the sixth to second order cumulant towards the lowest inspected energy. As expected the contribution of excited cluster states is most important at low energies $\sqrt{s_{NN}}<3.5$~GeV and becomes negligible at higher collision energies. Especially in light of the expected ultra-high precision data from CBM at FAIR, this new contribution is important to allow for a quantitative comparison with (potentially later available) lattice QCD or effective model results.
\end{abstract}




\end{frontmatter}




\section{Introduction}
\label{s:intro}
Heavy ion reactions are nowadays the most promising tool to explore the properties of Quantum-Chromo-Dynamics (QCD) at high densities and temperatures. In contrast to indirect studies, e.g. using gravitational waves from binary neutron star mergers \cite{LIGOScientific:2017vwq}, laboratory experiments at CERN, RHIC, and GSI allow for a rather precise selection of the created temperature and baryon density by varying the mass of the colliding system and the collision energy \cite{Sorensen:2023zkk}. Especially the RHIC beam energy scan (BES) program \cite{Odyniec:2019kfh} measured a multitude of collision energies in the range where the critical end point (CEP) of QCD phase diagram is expected \cite{Stephanov:2024xkn}. Various possible signatures for the observation of the CEP have been put forward in the literature, ranging from the analysis of HBT correlations \cite{Stephanov:1998dy,Lacey:2014wqa}, intermittency analysis \cite{Adhikary:2023rfj}, the formation of baryon droplets \cite{Herold:2013qda} to the analysis of (non-)Gaussian fluctuations of conserved quantities \cite{Stephanov:2008qz,HotQCD:2012fhj}. Especially the last point has emerged as most promising, because fluctuations (or more specifically, susceptibilities) can be extracted from first principle lattice QCD simulations and measured directly in experiment. Unfortunately, some caveats are still present that make the unique interpretation of the experimental data difficult, e.g. the acceptance cuts in experiments influence the fluctuation signal (generally, a large acceptance leads to vanishing fluctuations of conserved quantities, while a very small acceptance will always yield Poisson fluctuations) \cite{Nahrgang:2009dqc,Braun-Munzinger:2023gsd,Kuznietsov:2024xyn}, also the finite size and life-time of the system influences the observed fluctuations \cite{Nahrgang:2011mg}, further volume fluctuations may mask the physical fluctuations \cite{Skokov:2012ds,Holzmann:2024wyd}, or not all sources of fluctuation have been included in the (theoretical and/or experimental) analysis \cite{Sombun:2017bxi,Feckova:2015qza}. 

Here we report on a novel source of fluctuations, which was omitted in previous analyses, namely contributions from decays of excited nuclear cluster states. Up to now, standard baseline calculations for the fluctuation observable have omitted the contributions form the decay of high mass nuclear cluster states. It is well known that the decay of excited clusters gives a sizable contribution already to the average yields of deuterons, tritons, etc.~\cite{Vovchenko:2020dmv}. However, the contributions of these states to the fluctuations of conserved charges---or more specifically the proton number as the proxy for baryon number fluctuations---have not yet been considered.

While it is clear that these contributions are small at high collision energies ($\sqrt{s_{NN}}>5$~GeV) because the yield of clusters is already small, at lower collision energies this additional contribution becomes important. Especially, with the construction of the CBM experiment at FAIR, which will provide unprecedented reaction rates, the exploration of higher order cumulants becomes possible and improved baseline calculations are necessary. Only then precision tests of lattice QCD results at finite baryon density using susceptibilities are possible.

Here we employ the list of excited cluster states taken from \cite{Vovchenko:2020dmv,Tilley:1992zz,Tilley:2002vg} and include their decays into a thermal model analysis at various collision energies. The approach is detailed in the next Section. We present the results in Section \ref{s:calc_n_res} and conclude in Section~\ref{s:conc}.


\section{Formalism}
\label{s:form}

For our calculation we use the grand-canonical ensemble \cite{Nahrgang:2014fza,Tomasik:2021jfd}. 
The formalism closely follows our earlier work, so we review the main relations here and refer the reader to \cite{Tomasik:2021jfd} for more details. 

In a resonance gas treated grand-canonically, proton number fluctuates due to two reasons: 
Firstly, protons are exchanged with the heatbath. 
Secondly, resonance decays are random processes which produce protons with some probability. 
Excited nuclear clusters contribute to proton number fluctuations through their decays. 
Hence, for the sake of our calculation, technically we can treat those clusters similar to unstable resonances. 
Moreover, like for baryonic resonances, all included excited clusters produce at most one proton. 
Hence, the probability distribution for the number of protons resulting from decays of $N_R$ resonances or clusters of the same type is binomial. 

We will look at the cumulants of the (net-)proton number up to sixth order, which we can obtain from the cumulants of proton and antiproton numbers
\begin{equation}
\cum{(\delN_\netp)}{l} = \cum{(\delN_p)}{l} + (-1)^l \cum{(\delN_{\bar p})}{l}\,  ,
\label{e:cumppbar}
\end{equation}
where $l$ denotes the order of the cumulant. For nuclear collisions at lower energies the net-proton number is practically equal to the number of protons, though. 
In addition the mean and variance, we will use the standardized moments of the cumulants:
the skewness and the kurtosis
\begin{eqnarray}
M & = & \langle N \rangle \, ,\\
\sigma^2 & = & \cum{(\delN)}{2}\,  ,\\
S & = & \frac{\cum{(\delN)}{3}}{{\cum{(\delN)}{2}}^{\!\!\!3/2}}\,  , \\
\kappa & = & \frac{\cum{(\delN)}{4}}{{\cum{(\delN)}{2}}^{2}}\,  ,
\end{eqnarray}
as well as the hyperskewness and hyperkurtosis for the fifth and sixth order, respectively
\begin{eqnarray}
S^H& = & \frac{\cum{(\delN)}{5}}{{\cum{(\delN)}{2}}^{\!\!\!5/2}}\,  , \\
\kappa^H & = & \frac{\cum{(\delN)}{6}}{{\cum{(\delN)}{2}}^{3}}\,  .
\end{eqnarray}
Since the cumulants scale with the volume, which is experimentally poorly controlled, ratios are constructed where the volume cancels out 
\begin{align}
\nonumber
 \frac{\cum{(\delN)}{2}}{\cum{N}{}} & =   \frac{\sigma^2}{M}  & &  \nonumber \\
 \frac{\cum{(\delN)}{3}}{\cum{N}{}} & = \frac{S \sigma^3}{M}   & 
 \frac{\cum{(\delN)}{3}}{\cum{(\delN)}{2}} & = S\sigma
\nonumber  \\
\frac{\cum{(\delN)}{4}}{\cum{(\delN)}{2}} & = \kappa\sigma^2 & & 
\label{e:volind1} 
\end{align}
and for the fifth and sixth order
\begin{align}
\frac{\cum{(\delN)}{5}}{\cum{N}{}} & = \frac{S^H\sigma^5}{M}  & 
\frac{\cum{(\delN)}{5}}{\cum{(\delN)}{2}} & = S^H\sigma^3  \nonumber \\
\frac{\cum{(\delN)}{6}}{\cum{(\delN)}{2}} & =\kappa^H\sigma^4  \,  . &&
\label{e:volind2}
\end{align}

In calculation, cumulants are obtained by taking derivatives of the cumulant-generating function
\begin{equation}
\cum{(\delN)}{l} = \left . \frac{\mathrm{d}^l K(i\xi)}{\mathrm{d} (i\xi)^l}\right |_{\xi=0}\,  .
\label{e:cumdef}
\end{equation}
For protons from resonance decays, the cumulant-generating function reads \cite{Tomasik:2021jfd}
\begin{equation}
K(i\xi) = \sum_R \ln \left \{ \sum_{N_R=0}^\infty P_R(N_R) \left ( e^{i\xi}\pr + (1-\pr) \right )^{N_R} \right \}\,  .
\end{equation}
The sum over $R$ counts all resonance species and it also includes the excited clusters which decay into protons. 
The probability that resonance $R$ produces one proton is denoted $\pr$.
Formally, direct protons are included as resonances that decay to protons with probability $\pr=1$.
The inner sum runs over all numbers $N_R$ of resonance species $R$, and $P_R(N_R)$ is the probability of number $N_R$ to occur. 
In our calculation, those probabilities are taken from the statistical thermal model.  

From the cumulant-generating function, the first six cumulants of the proton number distributions are obtained \cite{Tomasik:2021jfd}
\begin{subequations}
\label{e:allpcums}
\begin{eqnarray}
\cum{N_p}{} & = & \sum_R \pr \cum{N_R}{}\,  \\
\cum{(\delN_p)}{2} & = & \sum_R \left [ \pr^2 \cum{(\delN_R)}{2} + \pr(1-\pr) \cum{N_R}{} \right ] \, , \\
\cum{(\delN_p)}{3} & = & \sum_R \Bigl [  \pr^3 \cum{(\delN_R)}{3}  + 3 \pr^2(1-\pr)  \cum{(\delN_R)}{2} 
 \nonumber \\ && \qquad {}
+ \pr(1-\pr)(1-2\pr)\cum{N_R}{} \Bigr ] \,  , \\
\cum{(\delN_p)}{4} & = & \sum_R \Bigl [
\pr^4\cum{(\delN_R)}{4} + 6\pr^3(1-\pr)\cum{(\delN_R)}{3}
\nonumber \\ && \qquad {}
 + \pr^2(1-\pr)(7-11\pr)\cum{(\delN_R)}{2} 
 \nonumber \\
 && \qquad   {}
+ \pr(1-\pr)(1-6\pr+6\pr^2)
\nonumber \\ && \qquad \quad {} \times
\cum{N_R}{} \Bigr ] \,  , \\
\cum{(\delN_p)}{5} & = & \sum_R \Bigl [
\pr^5\cum{(\delN_R)}{5} +10 \pr^4(1 - \pr)  \cum{(\delN_R)}{4} 
\nonumber \\ && \qquad {}
+ 5 \pr^3 (1-\pr)( 5-7\pr) \cum{(\delN_R)}{3}
\nonumber \\ 
&& \qquad {}
+5 \pr^2 (1-\pr)(10\pr^2 - 12\pr+3)
\nonumber \\ && \qquad \quad {} \times \cum{(\delN_R)}{2} 
\nonumber \\ 
&& \qquad {}
+  \pr(1 - \pr)  (1 - 2 \pr)  
\nonumber \\ && \qquad \quad {} \times 
(12\pr^2 - 12 \pr  +1)  \cum{N_R}{} \Bigr ]\, ,
\\
\cum{(\delN_p)}{6} & = & \sum_R \Bigl [
\pr^6\cum{(\delN_R)}{6} +15\pr^5(1-\pr)\cum{(\delN_R)}{5}
\nonumber \\ && \qquad {}
+5\pr^4(1-\pr)(13-17\pr)\cum{(\delN_R)}{4}
\nonumber \\ 
&& \qquad {}
+ 15\pr^3(1-\pr)(15\pr^2-20\pr+6)
\nonumber \\ && \qquad \quad {} \times
\cum{(\delN_R)}{3} 
\nonumber \\ 
&& \qquad {}
- \pr^2(1-\pr)
\nonumber \\ && \qquad \quad {} \times
(  274 \pr^3 -476 \pr^2  +239 \pr - 31  )
\nonumber \\ && \qquad \quad {} \times
\cum{(\delN_R)}{2}
\nonumber \\ 
&& \qquad {}
+ \pr(1-\pr)
\nonumber \\ && \qquad \quad {} \times
( 120 \pr^4 - 240 \pr^3 + 150 \pr^2  
\nonumber \\ && \qquad \qquad\qquad  {} 
- 30 \pr + 1)
\cum{N_R}{}  \Bigr ]\,  ,
\end{eqnarray}
\end{subequations}
where the cumulants of the resonance number distribution is obtained by taking derivatives of the  partition function for each resonance species $R$,
\begin{equation}
\label{e:Rcum}
\cum{(\delN_R)}{l} =\frac{\partial^l \ln {\cal Z}_R}{\partial (\mu_R/T)^l}\,  .
\end{equation}
Here, $\mu_R$ is the chemical potential of that species. For the partition function we assume a hadron resonance gas that consists of hadrons and resonances all with vanishing widths. The partition function for a single resonance species is then
\begin{equation}
\ln {\cal Z}_R =\frac{g_RV}{2\pi^2} m_R^2 T \sum_{j=1}^\infty \frac{(\mp 1)^{j-1}}{j^2} e^{j\mu_R/T} K_2\left ( \frac{jm_R}{T} \right )\, ,
\end{equation}
where $g_R$ is the spin degeneracy, $m_R$ is the mass of the resonance (or cluster), and $V$ is the volume of the whole system. 
By taking the derivatives, expressions for $\cum{(\delN_R)}{l}$ are obtained
\begin{eqnarray}
\cum{N_R}{} & = & \frac{g_RV}{2\pi^2} m_R^2 T 
\nonumber \\ &&  \times 
\sum_{j=1}^\infty \frac{(\mp 1)^{j-1}}{j} e^{j\mu_R/T} K_2\left ( \frac{jm_R}{T} \right )\, , 
\label{e:meanR}
\\
\cum{(\delN_R)}{l} & = & \frac{g_RV}{2\pi^2} m_R^2 T 
\nonumber \\ &&  \times
\sum_{j=1}^\infty (\mp 1)^{j-1}{j}^{l-2} e^{j\mu_R/T} K_2\left ( \frac{jm_R}{T} \right )\, .
\label{e:cumR}
\end{eqnarray}
Note that in the calculations of the cumulant ratios the volume cancels out, and so one does not need to specify it. 

We will calculate the cumulant ratios (\ref{e:volind1}) and (\ref{e:volind2}), where the numerators and denominators are determined through eqs.~(\ref{e:allpcums}). 
The cumulants of the resonance number distributions is inserted there from eqs.~(\ref{e:meanR}) and (\ref{e:cumR}).


\section{Calculation and results}
\label{s:calc_n_res}

In our calculation we include all resonance states that were included in \cite{Tomasik:2021jfd}. 
In addition to that, we put in as stable particles the ground states of nuclear clusters, including those of antinucleons, up to $A=4$. 
The excited cluster states are technically treated in the calculation as resonances that decay into protons with a given probability. 
Both stable and excited clusters are taken from Tables~I and II of \cite{Vovchenko:2020dmv}. 

The calculations are performed in chemical equilibrium given by temperature $T$ and baryochemical potential $\mu_B$. 
There are several parametrization of the collision energy dependence of $T$ and $\mu_B$ at the chemical freeze-out. 
Here, we adopt the functional forms used in \cite{Vovchenko:2015idt}
\begin{eqnarray}
\mu_B & = & \frac{d}{1+e\sqrt{s_{NN}}} 
\label{e:mub}\\
T & = & a - b\mu_B^2 - c\mu_B^4\, .
\label{e:T}
\end{eqnarray}
The values of the parameters are summarized in Table~\ref{t:coef}
\begin{table}
\begin{tabular}{ccccc}
\hline
$a$ [GeV] & $b$ [GeV$^{-1}$] & $c$ [GeV$^{-3}$] & $d$ [GeV] & $e$ [GeV$^{-1}]$ \\
  0.157  & 0.087  & 0.092  & 1.477 & 0.343\\
\hline
\end{tabular}
\caption{Values of coefficients for the determination of $T$ and $\mu_B$ at the chemical freeze-out.
}
\label{t:coef}
\end{table}

\begin{figure*}[t]
	\centering 
	\includegraphics[width=0.73\textwidth]{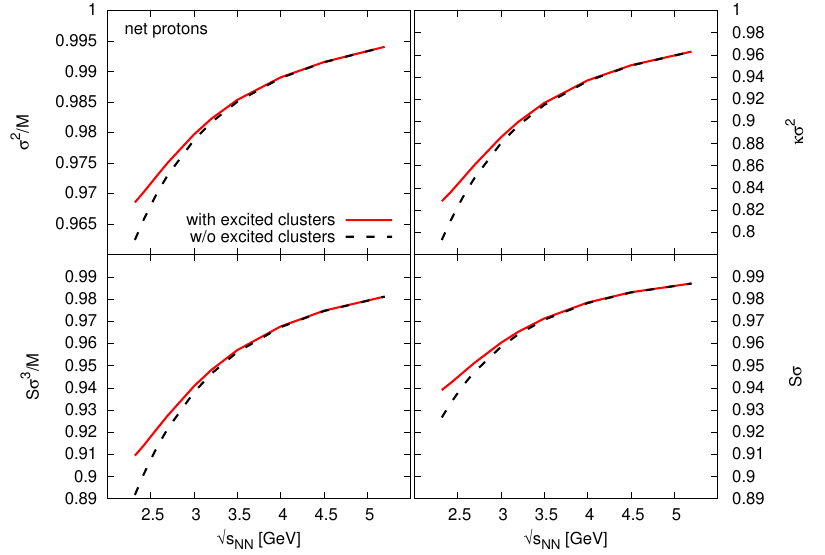}	
	\caption{Dependence of volume-independent ratios of the first four cumulants of the net-proton number distribution
	on the collision energy.
	The relation to the cumulant ratios is explained in Eqs.~(\ref{e:volind1}).} 
	\label{f:c2-4}
\end{figure*}

\begin{figure*}[t]
	\centering 
	\includegraphics[width=0.73\textwidth]{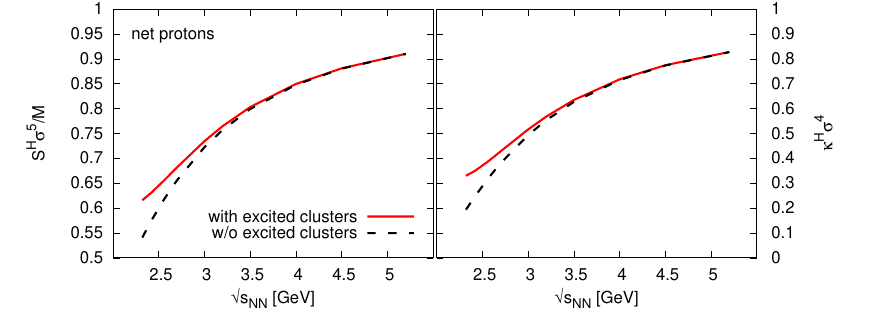}	
	\caption{Dependence of volume-independent ratios that include the fifth and sixth cumulants of the 
	net-proton number distribution
	on the collision energy.
	The relation to the cumulant ratios is explained in Eqs.~(\ref{e:volind2}).} 
	\label{f:c56}
\end{figure*}

We show the ratios of the first four cumulants in Fig.~\ref{f:c2-4}. 
The ratios that include hyperskewness and hyperkurtosis are depicted in Fig.~\ref{f:c56}.
In all cases, the shape of the obtained collision energy dependencies looks quite similar, although they are different in magnitude. Generally, the inclusion of the excited nuclear states increases the ratios and brings them closer to unity. 
Quantitatively, we find that the inclusion of the excited nuclear clusters yields corrections to all cumulant ratios, ranging from 1\% ($\cum{(\delN)}{2}/\cum{N}{}$) to 100\% ($\cum{(\delN)}{6}/\cum{(\delN)}{2}$) towards the lowest inspected energy.
The effect is particularly visible for $\sqrt{s_{NN}}$  below 3.5~GeV. 

\begin{figure}[t]
	\centering 
	\includegraphics[width=0.48\textwidth]{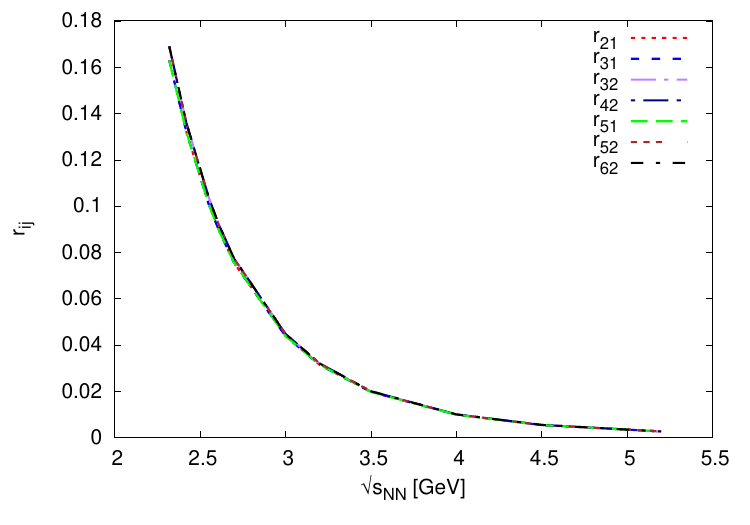}	
	\caption{The scaling variable $r_{ij}$, defined in Eqs.~(\ref{e:ri1}) and (\ref{e:rij}) as depending on collision energy, for various ratios of the cumulants. }
	\label{f:ratio}
\end{figure}

A closer inspection reveals that the excited nuclear states have the same relative influence to all orders of the cumulant ratios. 
In order to show this, we define a scaling variable 
\begin{equation}
\label{e:ri1}
r_{i1}   =  \frac{1 - \cum{(\delN)}{i}/\cum{N}{} (\mbox{with clusters})}{1 - \cum{(\delN)}{i}/\cum{N}{} (\mbox{w/o clusters})}\, ,
\end{equation}
and for $j>1$
\begin{equation}
r_{ij}  =  \frac{1 - \cum{(\delN)}{i}/\cum{(\delN)}{j} (\mbox{with clusters})}{1 - \cum{(\delN)}{i}/\cum{(\delN)}{j} (\mbox{w/o clusters})}\, ,
\label{e:rij}
\end{equation}
so that the numerator is determined in calculation where excited nuclear clusters are included, while they are left out in the denominator. This variable measures the relative distance that the cumulant ratio calculated with excited clusters gets closer to unity in comparison to the ratio calculated without clusters. The energy dependence of the ratio is shown in Fig.~\ref{f:ratio}.
We clearly see that all the investigated ratios collapse to one curve. 
To be more precise, there are actually two curves: one for the ratios with $\cum{N}{}$ in denominator, another for those with $\cum{(\delN)}{2}$. 
This shows that the effect of the excited nuclear states cuts up to one sixth from the distance of the cumulant ratio from unity, at collision energy 2.4~GeV. 

In absolute numbers, however, the effect is best seen in Figs.~\ref{f:c2-4} and \ref{f:c56}. 
While  ${\cum{(\delN)}{2}}/{\cum{N}{}}$  is close to 1 and increases by less than 1\% in absolute numbers, the ratio ${\cum{(\delN)}{6}}/{\cum{(\delN)}{2}}$ almost doubles. Hence, the effect gains on importance once higher-order cumulants are examined. 


\section{Conclusion}
\label{s:conc}
We have calculated the contribution of the decay products of excited nuclear cluster states to the event-by-event fluctuations of protons in the energy range from $\sqrt{s_{NN}}=2-5$~GeV. To this aim we assumed validity of the statistical model, with parameters adjusted to describe average multiplicities for central Au+Au reactions at each investigated collisions energy. We found that the inclusion of the excited nuclear clusters yields corrections to all cumulant ratios, ranging from 1\% ($\cum{(\delN)}{2}/\cum{N}{}$) to 100\% ($\cum{(\delN)}{6}/\cum{(\delN)}{2}$) towards the lowest inspected energy. As expected, the contribution of excited cluster states is most important at low energies $\sqrt{s_{NN}}<3.5$~GeV and becomes negligible at higher collision energies. Especially in light of the expected ultra-high precision data from CBM at FAIR, this new contribution is important to allow for a quantitative comparison with (potentially later available) lattice QCD or effective model results.


\section*{Acknowledgments}
BT thanks the Czech Science Foundation for the support under grant No.~22-25026S, 
and to  VEGA under grant No.~1/0521/22.

\bibliographystyle{elsarticle-num} 
\bibliography{p_fluct_exc.bib}






\end{document}